\documentclass[twocolumn,showpacs,showkeys,preprintsnumbers,amsmath,amssymb]{revtex4}

\usepackage{graphicx}
\usepackage{epsfig}
\usepackage{dcolumn}
\usepackage{bm}

\begin{document}

\title  
{ Perturbative expansion using variational phonon basis for Holstein model }
\author 
{
Jayita Chatterjee$^1$, A. N. Das$^1$ and P. Choudhury$^2$
}
\affiliation
{
$^1$ Saha Institute of Nuclear Physics \\
1/AF Bidhannagar, Calcutta 700 064, India\\
$^2$ Central Glass and Ceramic Research Institute \\
196 Raja Subodh Chandra Mullick Road, Calcutta 700 032, India\\
}
\begin{abstract}
 A simple variational displacement phonon basis, obtained through the
modified Lang-Firsov (MLF) transformation, is proposed to study the Holstein
model. This phonon basis contains only one variational parameter, but capable
of describing lattice distortions at distant sites from the charge carrier.
Perturbation method based on this MLF basis is employed to calculate the
 single-electron ground-state energy and the dispersion of the polaronic band.
The ground-state ($k$=0) energy
obtained up to the second-order perturbation within this approach agrees well
with the available numerical results for the entire range of coupling strength.


\end{abstract}

\maketitle

\vskip 1.5 cm

Self-localization of electrons in crystal lattices due to interaction 
with phonons constitutes one of the central problems in condensed matter 
physics. The simplest Holstein model \cite{Hol}, describing coupling 
of conduction 
electron with dispersionless phonons, has been studied in weak \cite{Mig}
as well as strong coupling \cite{LF} limits following either variational
\cite{TOY,Fein,DS} or perturbation methods \cite {Mar,Zoli,jayee}.
Numerical approaches including exact diagonalization (ED) 
techniques \cite{Mar,Wellein}, Quantum Monte Carlo calculations 
\cite {Korn}, 
Global-Local (GL) method \cite{Rom}, density-matrix-renormalization-group 
(DMRG) technique \cite{White} and recently 
developed ED technique with variational Hilbert space \cite{Trug} 
are quite successful in evaluating 
the ground-state properties including various correlation functions 
in the entire range of electron-phonon ($e$-ph) coupling.
The analytical methods, however, suffer from many limitations.  
Migdal approximation \cite{Mig} is valid for a weakly-coupled 
adiabatic system while 
Lang-Firsov (LF) canonical transformation \cite{LF} is suitable  
only in the strong-coupling limit. 
No common consensus regarding proper choice of the phonon 
basis and applicability of perturbation theory has yet been reached 
for the whole range of $e$-ph coupling. So far the perturbative 
method has been employed using the LF basis only and no perturbation
calculation based on the modified LF (MLF) basis is available. 
Recently, we \cite{jayee} investigated the convergence of 
the perturbation expansion on a two-site single electron 
Holstein model using different phonon bases obtained through LF, 
MLF and MLF with squeezing (MLFS) transformations. 
It has been observed that for a wide region of e-ph coupling the perturbation 
corrections within the MLF method are much smaller and the convergence 
is much better compared to the LF approach. 

Inspired with this result, we now attempt to apply the perturbation 
method based on a properly chosen MLF basis to the one-dimensional 
(infinite site) Holstein model and evaluate the perturbation
 corrections to the ground-state energy, wave function and the 
dispersion.
The Holstein Hamiltonian for a single electron reads as 
\begin{eqnarray}
H &=& \omega \sum_i  b_i^{\dag} b_i 
        + \epsilon \sum_{i} n_{i}
	- t \sum_{i , \delta_1} c_i^{\dag} c_{i + \delta_1} \nonumber\\
&+& \omega g  \sum_i (b_i^{\dag} + b_i) n_i 
\end{eqnarray} 
where $i$ =1 to $N$, denotes the site. $c_{i}$ ($c_{i}^{\dag}$)  
is the annihilation (creation) operator for the electron 
at site $i$ and $n_{i}$ (=$c_{i}^{\dag} c_{i}$) 
is the corresponding number operator, $g$ denotes the on-site $e$-ph 
coupling strength and $t$ is the usual hopping integral. $b_i$ and 
$b_{i}^{\dag}$ are the annihilation and creation operators, respectively, 
for the phonons corresponding to 
interatomic vibration at site $i$ and $\omega$ is the phonon frequency. 
$\delta_1$ runs over nearest-neighbor sites.  

The spread and depth of the polaron may be studied conveniently 
within the framework 
of the MLF transformation where the lattice deformations produced  
by a charge carrier at different neighboring sites are treated as 
variational parameters \cite{Fein,DS}. The MLF transformation on 
Eq. (1) gives $\tilde{H} = e^R H e^{-R}$\\ 
where \\ 
\begin{eqnarray}
 R  &=&  \sum_i n_i  [ \lambda_0  (b_i^{\dag} - b_i) 
   ~+~ \lambda_1 \sum_{\delta_1}(b_{i+ \delta_1}^{\dag} - 
     b_{i+ \delta_1}) \nonumber \\
  &+&  \lambda_2 \sum_{\delta_2}(b_{i + \delta_2} ^{\dag} - 
     b_{i + \delta_2}) 
~+...+ \lambda_l \sum_{\delta_l}(b_{i + \delta_l}
   ^{\dag} - b_{i+\delta_l}) +..] \nonumber  
\end{eqnarray} 
and $\lambda_0$, $\lambda_1$, $\lambda_2$, ... represent 
the variational lattice
deformations around the charge carrier produced at the same, 
first-, second-, etc. neighbouring sites, respectively. $\delta_{l}$
runs over the $l$ th nearest neighbour sites.
For LF transformation, $\lambda_0$=$g$
and $\lambda_1=\lambda_2= ... =\lambda_l=...=0$. 

The MLF transformed Hamiltonian is then obtained as 
\begin{eqnarray}
\tilde{H} &=& \omega \sum_i b_i^{\dag} b_i + \epsilon_p \sum_i n_i
+H_t + H_s \\
{\rm where} \nonumber \\
 H_t &=& - t \sum_{i, \delta_1} 
{\rm{exp}} (Y_i - Y_{i + \delta_1}) ~c_i^{\dag} c_{i + \delta_1}~, \nonumber \\
 H_s &=& \omega \sum_i (b_i^{\dag} + b_i) [(g - \lambda_0) n_i 
- \lambda_1 \sum_{\delta_1} n_{i + \delta_1} \nonumber \\
 &-& \lambda_2 \sum_{\delta_2} n_{i + \delta_2}
- \lambda_3 \sum_{\delta_3} n_{i + \delta_3} -... ]~,\nonumber \\
Y_i &=& \lambda_0 (b_i^{\dag} - b_i)
+ \lambda_1 \sum_{\delta_1} (b^{\dag}_{i + \delta_1} - b_{i + \delta_1})
\nonumber \\
&+& \lambda_2 \sum_{\delta_2} (b^{\dag}_{i + \delta_2} - b_{i + \delta_2}) 
+ \lambda_3 \sum_{\delta_3} (b^{\dag}_{i + \delta_3} - b_{i + \delta_3})+ .. ,
\nonumber \\
{\rm and} \nonumber \\
\epsilon_p &=& \epsilon - \omega~(2 g - \lambda_0)~\lambda_0 
+ 2 \omega~(\lambda_1^2 + \lambda_2^2 + \lambda_3^2~+... ) \nonumber
\end{eqnarray}
is the polaron self-energy.

 For the perturbation calculation following Marsiglio \cite{Mar} 
we choose the basis set for the one-electron case as \\ 
\begin{equation}
 |\psi_l{\{n_i\}}\rangle = \frac{1}{\sqrt{N}} c_{l}^{\dag} 
|0\rangle_e  |n_1~n_2~..~n_N
\rangle_{ph}  
\end{equation}
which represents a state with $l$th site occupied by the electron and $n_1, n_2, ... , n_N$ being 
the phonon numbers at sites 1, 2, ... , N, respectively in the MLF
displaced phonon basis. $l$ takes values 
from 1 to $N$ while $n_1, n_2$, ... from zero to infinity covering the 
whole Hilbert space. 
$H_t$ and $H_s$ are non-diagonal and the remaining 
part of the Hamiltonian is diagonal in this representation. The 
unperturbed energy of the state, described in Eq. (3), is given by 
$E^{(0)}_{\{n_i\}} = \epsilon_p + n_{T}~ \omega$ where $n_{T}(= \sum_{i} n_i$)
represents the total number of phonons 
in that state. The ground state corresponds to $n_{T}=0$ and has 
$N$-fold degeneracy since the electron can occupy any one of the $N$ 
sites. This degeneracy is lifted by considering the first-order 
energy correction which results in a polaronic band with energy 
$E_k = \epsilon_p - 2 t_e \cos(ka)$ 
where $t_e= t \exp(-\{ (\lambda_0 - \lambda_1)^2 +
(\lambda_1 - \lambda_2)^2+ (\lambda_2 - \lambda_3)^2+..  \})$.

 The eigenstates are now given by 
\begin{equation}
|\psi_k\rangle= \frac{1}{\sqrt N}\sum_{l}
\exp{(i\vec{k}.\vec{R}_l)}~c_{l}^{\dag} |0\rangle_e  |0\rangle_{ph} 
\end{equation}
and the unperturbed ground state $|\psi_G^{(0)}\rangle$ corresponds to $k=0$.

 For perturbation calculation it is very tedious and inconvenient
to deal with a phonon basis containing too many variational parameters.
Previously we \cite{jayee} observed that a choice of variational 
parameters
which minimizes the ground-state energy, obtained within
MLF and zero-phonon averaging (ZPA) (i.e., from 
$_{ph}\langle 0 |\tilde{H}| 0\rangle_{ph}$) leads to a phonon basis 
where the convergence of the perturbation series is satisfactory. Here
minimization of $E_{k=0} = \epsilon_p - 2 t_e $ (equivalent to the 
ground-state
energy obtained by ZPA 
of $\tilde{H}$) with respect to parameters
($\lambda_0$, $\lambda_1$, ...,$\lambda_l$)
gives the general recursion relation between the variational parameters 
\begin{eqnarray}
\frac{\lambda_{l-1}}{\lambda_l}= \frac{\lambda_{l-2} -2\lambda_{l-1}
+\lambda_l} 
{\lambda_{l-1} -2\lambda_{l}+\lambda_{l+1}} .
\end{eqnarray}
which is satisfied by a simple choice for the parameters
\begin{eqnarray}
\frac{\lambda_1}{\lambda_0}= \frac{\lambda_2}{\lambda_1}
= \frac{\lambda_3}{\lambda_2}=
... = \frac{\lambda_l}{\lambda_{l-1}} = ... = r.
\end{eqnarray}
A sum rule is also followed by $\lambda_l$'s; for one-dimensional chain 
this is given by 
\begin{eqnarray}
\lambda_0 + 2\lambda_1 + 2\lambda_2 + ... + 2\lambda_l + ... = g. 
\end{eqnarray}
Eq. (7) is a consequence of the fact that the in-phase phonon mode
(zero momentum)  being a sum of lattice displacement operators 
at different sites ($\sum_{i} (b_i+b_i^{\dag})$) 
couples only with the total number of electrons, which is 
a constant of motion.  
Eqs. (6) and (7) allow one to express all $\lambda_l$'s in terms 
of a single variational parameter $r$. Thus, the obtained MLF phonon 
basis, 
though being capable of considering lattice distortions at all sites 
of the infinite chain, contains only one variational parameter.

 The polaron self-energy may then be written in terms of $r$ as 
$\epsilon_p = \epsilon - \omega g^{2}\frac{(1-r)
(1+4r+r^{2})}{(1+r)^{3}}$ and the effective hopping
$t_e= t \exp(-g^{2} x^{3})$ where $x=\frac{1-r}{1+r}$.

The expression for the off-diagonal matrix element between the 
ground state and an excited state is given by 
\begin{eqnarray}
H_{l\{n\},G}^{'}&=& \langle \psi_l {\{n_i\}}| (H_t+ 
H_s)  |\psi_G^{(0)} \rangle \nonumber \\
&=& - \frac{t_e ~y^{n_T} }{\sqrt{N}} 
\left[\frac{ [(-1)^{n_{L}} ~r^{A} +(-1)^{n_{R}} ~r^{B} ] } 
{\sqrt{(...n_{l - 1}!~n_l ! ~ n_{l + 1}!....)}} ~\right]
\nonumber \\  
~&+&~\frac{g \omega}{\sqrt{N}} [(1-x) \delta_{n_l,1} -x[r 
(\delta_{n_{l - 1},1}+ \delta_{n_{l + 1},1} )   \nonumber \\
&+&  r^{2}(\delta_{n_{l - 2},1}+ \delta_{n_{l + 2},1} ) +
... ] ] \delta_{n_T,1}
\end{eqnarray}
\begin{eqnarray}
{\rm where} ~ y &=& g x (1-r), \nonumber\\
n_L &=&  n_{l-1}+n_{l-2}+n_{l-3}+... , \nonumber\\ 
n_R &=& n_{l+1}+n_{l+2}+n_{l+3}+... ,\nonumber \\
A &=& n_{l+1}+n_{l-2} + 2(n_{l+2}+n_{l-3}) \nonumber\\
&+& 3(n_{l+3}+n_{l-4})+.. ,\nonumber\\ 
B &=& n_{l-1}+n_{l+2} + 2(n_{l-2}+n_{l+3}) \nonumber\\
&+& 3(n_{l-3}+n_{l+4})+... .
\end{eqnarray}

 The second-order correction to the ground-state energy is given by
\begin{eqnarray}
E_0^{(2)} &=& -\frac{2 g^2 }{\omega} \left[ \omega r -t_e(1-r)^2 \right]^2 
\frac{(3+r)} {(1+r)^3}   \\   
&-&\frac{1}{N}\sum_l \sum_{n_T \ge 2} \frac{ t_e^2 y^{2 n_T} }{( n_T \omega)} 
\frac{ [(-1)^{n_L} ~r^A +(-1)^{n_R} ~r^B]^2 }{(...n_{l-1}!~n_l ! 
~ n_{l+1}!....)} \nonumber
\end{eqnarray}
and the first-order correction to the ground-state wave function is 
obtained as,  
\begin{eqnarray}  
|\psi_G^{(1)}\rangle = -\frac{1}{\omega \sqrt{N}} 
\sum_{l=1}^N c_{l}^{\dag} |0 \rangle_e 
[ \{\omega r - t_e (1-r)^2  \}  \nonumber\\ 
\sum_m [\frac{2g}{1+r}  \delta_{m,l}
- \sum_p  g x r^{p-1} \delta_{m,l+p}]~ |0 0 ..0~1_{m}~0..\rangle_{ph} ]  
\nonumber \\
-\frac{1}{\omega \sqrt{N}} \sum_{l=1}^N c_{l}^{\dag} |0 \rangle_e 
 ~\sum_{n_T \ge 2} \frac{H^{'}_{l\{n\},G}}
{~(n_T )}
 |n_1 ~n_2~...n_N\rangle_{ph}  
\end{eqnarray}

 Now one has to make a proper choice of $r$.  
In addition to Eq. (6) the minimization of $E_{k=0}$ yields a condition 
$\omega r = t_e(1-r)^2$, which gives 
\begin{eqnarray}
r=\frac {(2 t_e+ \omega) - \sqrt{4 t_e \omega + \omega^{2}} }
{ 2 t_e } \nonumber 
\end{eqnarray}
It is interesting to note that for this choice of $r$ the 
off-diagonal matrix element between the ground state and any excited 
state with a single phonon ($n_T=1$) becomes zero and this is consistent
with our previous studies \cite{jayee}. Correspondingly,    
the first term in Eq. (10) as well as in Eq. (11) vanishes.
\begin{center}
\begin{figure}[h]
\epsfxsize=3.4in
\epsffile{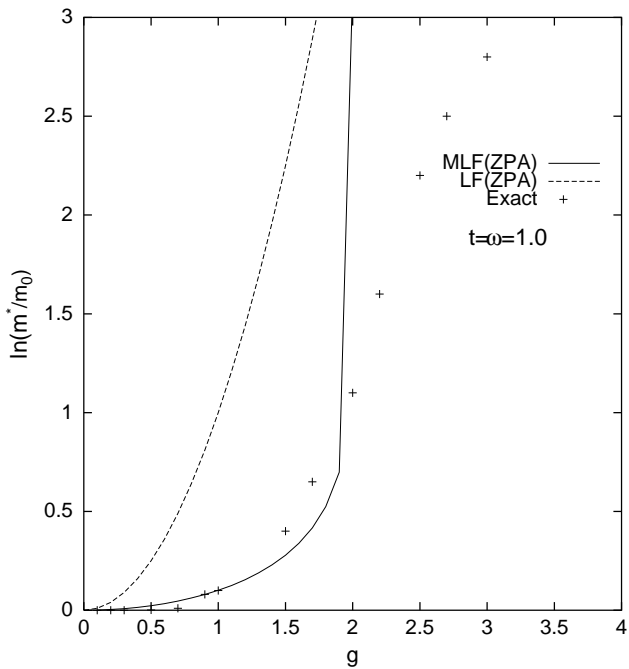}
\vskip 0.5 cm
\caption{
\label{figure1}
Plot of logarithm of the effective mass $m^*/m_0$ as function of $g$
for $t=\omega=1.0$ within MLF and LF approach using the unperturbed
ground-state wave function. `$+$' shows the results
of Ref [14].
}
\end{figure}
\end{center}
 For real materials of interest
the Hamiltonian may involve other interactions in addition to the 
Holstein-type $e$-ph interaction and the perturbation calculations 
for such complicated systems may not be possible. The common practice in
such cases is to make LF or MLF transformation followed by ZPA.
It may be mentioned that ZPA is equivalent to the zeroth (first) order
of perturbation as far as the wave function (energy) is concerned. The
success of the ZPA depends on the choice of the phonon basis.      
One of the advantages of the MLF phonon basis over LF basis is that
the second-order perturbation correction ($E_0^{(2)}$) within MLF 
basis is much smaller than that within LF basis except in the 
strong-coupling regime where both the bases become equivalent. 
For instance, for $\omega= t= 1$, $E_0^{(2)}(MLF) = -0.01746$ and $-0.1849$ 
while $E_0^{(2)}(LF)= -1.3538$ and $-0.4801$ for $g= 1.0$ and 1.7,
respectively.

 A comparison of the effective mass of the polaron obtained within the MLF 
and LF with ZPA may be helpful.
We have calculated the effective mass of the polaron using the 
standard formula 
$\frac{m_0}{m^*}=\frac{1}{2 t} \frac{\delta^{2} E(k)}{\delta k^{2}} |_{k=0}$ 
(where m$_0=\frac{1}{2 t}$) within the ZPA of LF and MLF Hamiltonian.
These results along with the exact numerical result of Ref. \cite{Trug}
are presented in Fig. 1. The proximity of the MLF results 
with the exact one indicates that the MLF basis is a better choice. 
\begin{center}
\begin{figure}[h]
\epsfxsize=3.5in
\epsffile{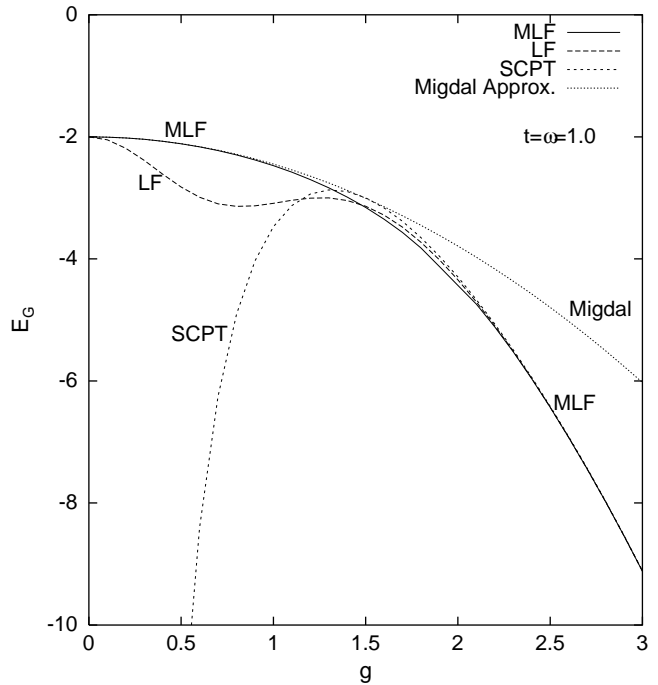}
\vskip 0.5 cm
\caption{
\label{figure2}
Plot of the ground-state energy (in unit of $t$) obtained by
considering perturbation 
corrections up to second-order within MLF, LF method and also within 
the weak-coupling (self-consistent Migdal approximation) and 
strong-coupling perturbation expansion (SCPT) obtained from Ref. [7] 
as a function of $g$ for $t=\omega=1.0$. 
}
\end{figure}
\end{center}
 
 In Fig. 2 we have shown the ground-state energy calculated
 up to the second-order perturbation correction within 
the LF (corresponds to $r=0$) and the MLF phonon basis. Energies 
calculated from the weak coupling (self-consistent Migdal approximation) 
and strong coupling perturbation limit, obtained from (Eqs. (7) and 
(C.17) of) Ref. \cite{Mar}, are also plotted. From comparison with the 
GL results \cite{Rom} we find that the MLF energy almost coincides 
with the exact energy in the entire regime of $g$. 
The weak-coupling method
fails for large $g$ whereas the strong-coupling perturbation expansion 
of Ref. \cite{Mar} breaks down and the LF perturbation result deviates 
much from the exact energy for low $g$ ($< 1.4$).

 To find out the energy dispersion within the MLF method
the energy $E_k$ has been calculated up to the second-order 
perturbation.
The second-order 
perturbation correction to $E_k$ within the MLF approach is 
obtained as 
\begin{eqnarray}
E_k^{(2)} &=& -\frac{2 g^2 }{\omega} 
\left[ \frac{(3+r)} {(1+r)^3} (\omega r -t_e(1-r)^2 {\rm cos}(ka)~)^2 
\right.\nonumber\\
&+&\left. \frac{(1-r)^3}{(1+r)} t_e^2~ {\rm sin}^2 (ka) \right]   \\   
&-&\frac{1}{N}\sum_l \sum_{n_T \ge 2} \frac{ t_e^2 y^{2 n_T} }{( n_T \omega)} 
\frac{(X^2+Y^2+2XY {\rm cos}(2ka))}{(...n_{l-1}!~n_l ! 
~ n_{l+1}!....)} \nonumber
\end{eqnarray}
where $X= (-1)^{n_L} r^A$ and $Y= (-1)^{n_R} r^B$, $n_L$, $n_R$, 
and $A$ and $B$ are defined in Eq. (9).

 The corresponding expression for $E_k^{(2)}$ within the LF approach is 
\begin{eqnarray}
E_{k}^{(2)} &=& -\frac{2 t_{e}^{2} }{\omega} 
 \left[\sum_{n=1}^{\infty}\frac{2}{n}~ \frac{g^{2n}}{n!} {\rm cos}^2(k a) 
\right.\nonumber\\
          &+& \left.\sum_{n=0}^{\infty} 
\sum_{m=1}^{\infty} \frac{g^{2 (n+m)}}{( n+m)~ n!~ m!} \right]
\end{eqnarray}
where $t_e= t~ e^{-g^2}$.  

We have calculated $E_k^{(2)}$, hence $E_k$ up to the second-order 
perturbation within 
the LF and MLF approach. The second-order correction $E_k^{(2)}$ 
and the ratio $E_k^{(2)}/E_k$ for different $k$ values are listed 
in Table I for several values of g for both the MLF and the LF approach 
for $t=\omega=1$.  

 The following features are evident from our study:
(i) The second-order MLF corrections are appreciably smaller than 
the corresponding LF corrections for low values of $g$ and $k$. 
(ii) For small and intermediate values of $g$ (e.g. $g=0.5$ and $1.0$)
the MLF correction increases with increasing $k$, however the MLF 
corrections are less than the corresponding LF values in the 
entire region of $k$ except at $ka = \pi/2$ where both the MLF and LF
corrections are same. In fact, for $ka = \pi/2$ the value of $r$
becomes zero within the present MLF approach so the MLF basis
reduces to the LF basis for $ka = \pi/2$. 
\begin{center}
\begin{table}
\caption{$ka/\pi$, second-order correction energy ($E_{k}^{(2)}$, in unit
of $t$) and realtive second-order correction ($E_{k}^{(2)}/E_{k}$)
for MLF and LF method respectively, (for
$g=0.5,~ 1$ and $2.0$, $t=\omega=1.0$) for the ground state of 
many-site one polaron problem.}
\begin{tabular}{cccccc}
\hline
& \multicolumn{2}{c}{MLF} &&\multicolumn{2}{c}{LF} \\ 

$ka/\pi$ & $E_k^{(2)}$ &$E_k^{(2)}/E_{k}$ 
&&$E_k^{(2)}$& $E_k^{(2)}/E_{k}$\\
\hline\\
{g = 0.5}& & & & &\\
\cline{1-1}
 0.00  &  -0.0010180  &  0.000482  & &  -1.014952  &  0.359587 \\
 0.10  &  -0.0094835  &  0.004680  & &  -0.953203  &  0.355067 \\
 0.20  &  -0.0364071  &  0.020493  & &  -0.791541  &  0.343899 \\
 0.30  &  -0.0881567  &  0.062911  & &  -0.591716  &  0.336728 \\
 0.40  &  -0.1841571  &  0.189624  & &  -0.430054  &  0.370296 \\
 0.50  &  -0.3683048  &  0.595669  & &  -0.368305  &  0.595669 \\
\hline \\
{g = 1.0}&~ &~ & & & \\
\cline{1-1}
 0.00  &  -0.0174239  &  0.007045  & &  -1.353833  &  0.438192 \\
 0.10  &  -0.0501885  &  0.020757  & &  -1.285706  &  0.430657 \\
 0.20  &  -0.1526451  &  0.067293  & &  -1.107347  &  0.409736 \\
 0.30  &  -0.3373983  &  0.162439  & &  -0.886884  &  0.382384 \\
 0.40  &  -0.5934159  &  0.310214  & &  -0.708525  &  0.365995 \\
 0.50  &  -0.6403983  &  0.390392  & &  -0.640398  &  0.390392 \\
\hline \\
{g = 2.0}&~ &~& & & \\
\cline{1-1}
 0.00  &  -0.3833072  &  0.086432  & &  -0.305533  &  0.070364 \\
 0.10  &  -0.3700764  &  0.083773  & &  -0.303269  &  0.069908 \\
 0.20  &  -0.3426137  &  0.078217  & &  -0.297343  &  0.068718 \\
 0.30  &  -0.3159575  &  0.072782  & &  -0.290017  &  0.067265 \\
 0.40  &  -0.3020421  &  0.070014  & &  -0.284090  &  0.066138 \\
 0.50  &  -0.2814511  &  0.065737  & &  -0.281826  &  0.065819 \\
\hline \\
\end{tabular}
\end{table}
\end{center}
(iii) The second-order perturbation correction 
within the LF method is symmetric around $ka = \pi/2$
as evident from Eq. (13). It is also minimum for $ka = \pi/2$.
(iv) For small and intermediate values of $g$ 
the second-order LF corrections are quite high. As a result $E_k$ up to the 
second-order correction within the LF may not be an accurate estimate
of $E_k$. Nevertheless we have cited the LF values for comparison with
the MLF values.
(v) In a range $1.9<g<2.2$ (for $t=\omega=1$) the LF second-order corrections
are slightly less than the corresponding MLF corrections.
For higher values of $g$ the MLF phonon basis 
reduces to the LF phonon basis and the results by these two methods
become equivalent.
(vi) With increasing $g$ the second-order LF correction decreases 
considerably and becomes more and more $k$-independent.     
$E_k^{(2)}/E_k$ within LF is around 0.027 for $g=2.5$, 
for the whole region of $k$.

 In Fig. 3a we plot $E_k$ vs. $ka/\pi$ for $g=0.5$ and $1.0$ for the 
MLF approach and in Fig. 3b a comparison with LF result and the
numerically exact results \cite{Rom,White,Trug}
is shown for $g=1$. The MLF result is close to the 
numerically exact results only for low $k$ but deviates considerably
from the exact curve for higher values of $k$.
In this region the second-order corrections are increasingly
higher implying that higher-order corrections are to be included to obtain 
accurate results. The LF result is much less satisfactory 
than the MLF result for $g=1$.    
The MLF perturbation method predicts a flat band (FIG. 3(b)) in the
region $0.7 \pi \le ka \le \pi$ which is consistent qualitatively
with the exact result \cite{Rom,Trug} for $g=1.0$.
However, the values of $E_{k}$ predicted by the
MLF in this region is found to be not consistent with the exact results.

 For large $g$ the LF second-order corrections are small in the entire region
of $k$, hence the dispersion given by the LF up to the second-order should
be satisfactory. However, no numerical result for $E_k$ for large $g$ is
available to us for a comparison.
\begin{center}
\begin{figure}
\epsfxsize=2.55in
\epsffile{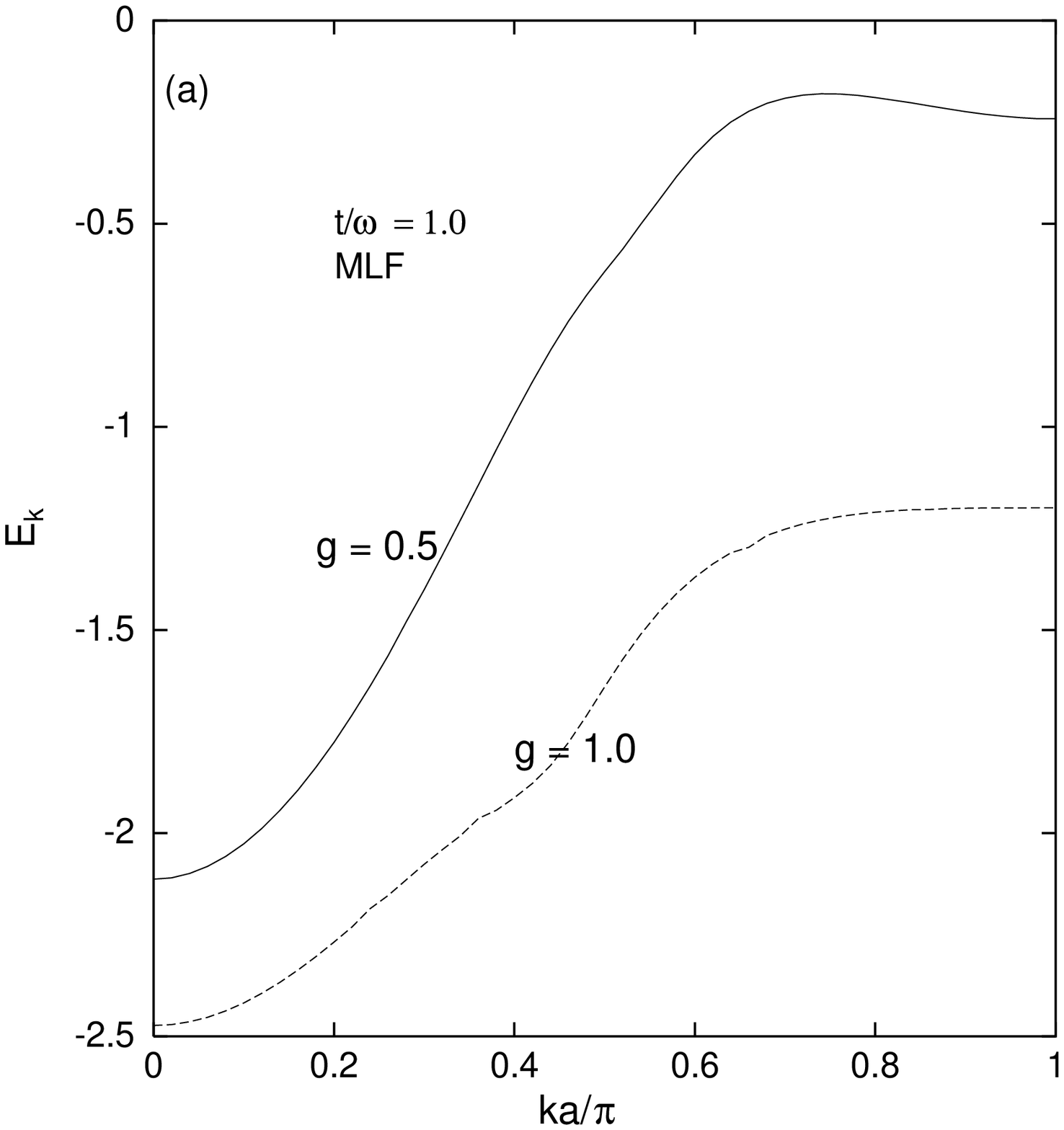}
\vskip 0.5 cm
\end{figure}
\end{center}
\begin{center}
\begin{figure} 
\epsfxsize=2.55in
\epsffile{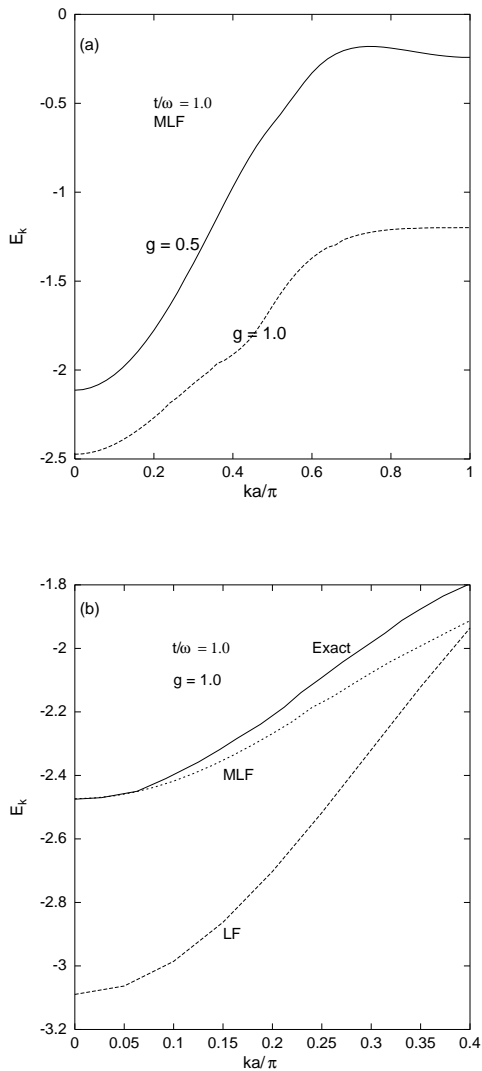}
\vskip 0.5 cm
\caption{
\label{figure3}
Plot of  $E_k$ (in unit of $t$) vs $ka/\pi$ for $t=\omega=1.0$:
(a) Within the MLF method for $g$ = 0.5 and 1.0.
(b) Within the MLF and LF methods for $g=1.0$; the solid line shows
the exact results from Ref.[14]. 
}
\end{figure}
\end{center}

 In summary, in this paper we have proposed a simple MLF 
phonon basis  containing a single variational parameter but capable
of describing lattice distortions at distant sites from the charge carrier.
For low and intermediate values of $g$ the second-order perturbation
correction to the energy within 
this approach is much smaller than that corresponding to the LF approach. 
It appears that the ZPA with this MLF basis gives much more
accurate results than that of 
ZPA with the LF basis. One representative calculation of effective mass is
presented which corroborates this feature. The ground-state ($k=0$) energy,
calculated up to the second-order  
within the MLF approach agrees well with exact results for the 
entire range of coupling strength. For $g\le1.0$, the second-order
correction within the MLF is small for low $k$ values and in this region the values of
$E_k$ (band dispersion) predicted by the MLF is consistent with the
numerically exact result. For large $g$ the LF perturbation corrections are small and the
band dispersion predicted by the LF method is expected to be satisfactory.
Computation of higher-order corrections would shed light on this issue.

Electronic address for correspondence: moon@cmp.saha.ernet.in

This work is partly supported by the Project No. SP/S2/M-62/96,
sponsered by Department of Science and Technology, India.

\end{document}